\documentclass[aps,prl,preprint,tightenlines,superscriptaddress,showpacs,byrevtex,floatfix]{revtex4}
\usepackage{amsmath2000}
\usepackage{graphicx}
\usepackage{dcolumn}
\usepackage{bm}

\begin{document}

\vspace*{-3\baselineskip}
\resizebox{!}{3cm}{\includegraphics{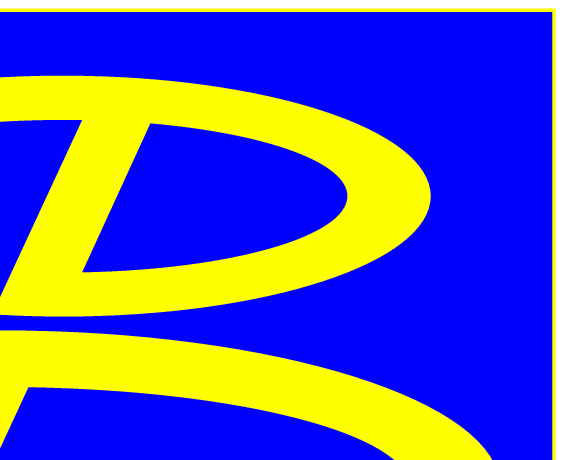}}

\preprint{KEK Preprint 2002-86}
\preprint{Belle Preprint 2002-30}

\title{Improved Measurement of Mixing-induced
{\boldmath $CP$} Violation in the Neutral {\boldmath $B$} Meson System}
%
\affiliation{Aomori University, Aomori}
\affiliation{Budker Institute of Nuclear Physics, Novosibirsk}
\affiliation{Chiba University, Chiba}
\affiliation{Chuo University, Tokyo}
\affiliation{University of Cincinnati, Cincinnati OH}
\affiliation{University of Frankfurt, Frankfurt}
\affiliation{University of Hawaii, Honolulu HI}
\affiliation{High Energy Accelerator Research Organization (KEK), Tsukuba}
\affiliation{Hiroshima Institute of Technology, Hiroshima}
\affiliation{Institute of High Energy Physics, Chinese Academy of Sciences, Beijing}
\affiliation{Institute of High Energy Physics, Vienna}
\affiliation{Institute for Theoretical and Experimental Physics, Moscow}
\affiliation{J. Stefan Institute, Ljubljana}
\affiliation{Korea University, Seoul}
\affiliation{Kyoto University, Kyoto}
\affiliation{Kyungpook National University, Taegu}
\affiliation{Institut de Physique des Hautes \'Energies, Universit\'e de Lausanne, Lausanne}
\affiliation{University of Ljubljana, Ljubljana}
\affiliation{University of Maribor, Maribor}
\affiliation{University of Melbourne, Victoria}
\affiliation{Nagoya University, Nagoya}
\affiliation{Nara Women's University, Nara}
\affiliation{National Kaohsiung Normal University, Kaohsiung}
\affiliation{National Lien-Ho Institute of Technology, Miao Li}
\affiliation{National Taiwan University, Taipei}
\affiliation{H. Niewodniczanski Institute of Nuclear Physics, Krakow}
\affiliation{Nihon Dental College, Niigata}
\affiliation{Niigata University, Niigata}
\affiliation{Osaka City University, Osaka}
\affiliation{Osaka University, Osaka}
\affiliation{Panjab University, Chandigarh}
\affiliation{Princeton University, Princeton NJ}
\affiliation{RIKEN BNL Research Center, Brookhaven NY}
\affiliation{Saga University, Saga}
\affiliation{University of Science and Technology of China, Hefei}
\affiliation{Seoul National University, Seoul}
\affiliation{Sungkyunkwan University, Suwon}
\affiliation{University of Sydney, Sydney NSW}
\affiliation{Tata Institute of Fundamental Research, Bombay}
\affiliation{Toho University, Funabashi}
\affiliation{Tohoku Gakuin University, Tagajo}
\affiliation{Tohoku University, Sendai}
\affiliation{University of Tokyo, Tokyo}
\affiliation{Tokyo Institute of Technology, Tokyo}
\affiliation{Tokyo Metropolitan University, Tokyo}
\affiliation{Tokyo University of Agriculture and Technology, Tokyo}
\affiliation{Toyama National College of Maritime Technology, Toyama}
\affiliation{University of Tsukuba, Tsukuba}
\affiliation{Utkal University, Bhubaneswer}
\affiliation{Virginia Polytechnic Institute and State University, Blacksburg VA}
\affiliation{Yokkaichi University, Yokkaichi}
\affiliation{Yonsei University, Seoul}
  \author{K.~Abe}\affiliation{High Energy Accelerator Research Organization (KEK), Tsukuba} 
  \author{K.~Abe}\affiliation{Tohoku Gakuin University, Tagajo} 
  \author{T.~Abe}\affiliation{Tohoku University, Sendai} 
  \author{I.~Adachi}\affiliation{High Energy Accelerator Research Organization (KEK), Tsukuba} 
  \author{H.~Aihara}\affiliation{University of Tokyo, Tokyo} 
  \author{M.~Akatsu}\affiliation{Nagoya University, Nagoya} 
  \author{Y.~Asano}\affiliation{University of Tsukuba, Tsukuba} 
  \author{T.~Aso}\affiliation{Toyama National College of Maritime Technology, Toyama} 
  \author{V.~Aulchenko}\affiliation{Budker Institute of Nuclear Physics, Novosibirsk} 
  \author{T.~Aushev}\affiliation{Institute for Theoretical and Experimental Physics, Moscow} 
% ...
\collaboration{The Belle Collaboration}

\date{\today}

\begin{abstract}
We present an improved measurement of the standard model $CP$ violation
parameter $\sinbb$ (also known as $\sin2\beta$) based on a sample
of $85 \times 10^6$ $B\overline{B}$ pairs
collected at the $\Upsilon(4S)$ resonance
with the Belle detector at the KEKB asymmetric-energy $e^+e^-$ collider.
One neutral $B$ meson is reconstructed in a $\jpsi\ks$, $\psi(2S)\ks$,
$\chi_{c1}\ks$, $\eta_c\ks$, $\jpsi\kstarz$, or $\jpsi\kl$ $CP$-eigenstate
decay channel and the flavor of the accompanying $B$ meson is
identified from its decay products.
From the asymmetry in the distribution of the time interval
between the two $B$ meson decay points, we obtain
$\sinbb = 0.719 \pm 0.074 \textrm{(stat)} \pm 0.035 \textrm{(syst)}$.
\end{abstract}


\maketitle

In the standard model (SM), $CP$ violation arises from an
irreducible complex phase in the weak interaction quark-mixing matrix
[Cabibbo-Kobayashi-Maskawa (CKM) matrix]~\cite{bib:ckm}.
In particular, the SM predicts a $CP$-violating asymmetry
in the time-dependent rates for $\bz$ and $\bzb$ decays
to a common $CP$ eigenstate $\fcp$,
where the transition is dominated by the $b \to c\overline{c}s$ process,
with negligible corrections from strong interactions~\cite{bib:sanda}:
\begin{equation}
  A(t) \equiv \frac{\Gamma(\bzb \to \fcp) - \Gamma(\bz \to \fcp)}
  {\Gamma(\bzb \to \fcp) + \Gamma(\bz \to \fcp)}
  = -\xi_f \sinbb \sin(\dM t),
\end{equation}
where $\Gamma(\bz,\bzb \to \fcp)$ is the rate for $\bz$ or $\bzb$
to $\fcp$ at a proper time $t$ after production,
$\xi_f$ is the $CP$ eigenvalue of $\fcp$,
$\dM$ is the mass difference between the two $\bz$ mass eigenstates,
and $\phi_1$ is one of the three interior angles of the CKM unitarity triangle,
defined as $\phi_1 \equiv \pi - \arg(V_{tb}^*V_{td}/V_{cb}^*V_{cd})$.
Non-zero values for $\sinbb$ have been reported
by the Belle and BaBar groups~\cite{bib:cpv,bib:babar}.

Belle's published measurement of $\sinbb$ is based on
a 29.1~fb$^{-1}$ data sample containing $31.3 \times 10^{6}$
$B\overline{B}$ pairs produced at the $\Upsilon(4S)$ resonance.
In this paper, we report an improved measurement that uses
$85 \times 10^{6}$ $B\overline{B}$ pairs (78~fb$^{-1}$).
Two important changes exist in the analysis
with respect to the published result~\cite{bib:cpv};
we apply a new track reconstruction algorithm that provides better performance
and a new proper-time interval resolution function~\cite{bib:blife}
that reduces systematic uncertainties in $\sinbb$.
The data were collected with the Belle detector~\cite{bib:belle}
at the KEKB  asymmetric collider~\cite{bib:kekb},
which collides 8.0~GeV $e^-$ on 3.5~GeV $e^+$
at a small ($\pm 11$~mrad) crossing angle.
We use events where one of the $B$ mesons decays to $\fcp$ at time $\tcp$,
and the other decays to a self-tagging state $\ftag$,
which distinguishes $\bz$ from $\bzb$, at time $\ttag$.
The $CP$ violation manifests itself as an asymmetry $A(\Dt)$,
where $\Dt$ is the proper time interval
between the two decays: $\Dt \equiv \tcp - \ttag$.
At KEKB, the $\Upsilon(4S)$ resonance is produced
with a boost of $\beta\gamma = 0.425$ nearly along the $z$ axis
defined as anti-parallel to the positron beam direction,
and $\Dt$ can be determined as $\Dt \simeq \Dz/(\beta\gamma)c$,
where $\Dz$ is the $z$ distance between the $\fcp$ and $\ftag$
decay vertices, $\Dz \equiv \zcp - \ztag$.
The average value of $\Dz$ is approximately 200~$\mu$m.

The Belle detector~\cite{bib:belle} is a large-solid-angle spectrometer
that includes a silicon vertex detector (SVD),
a central drift chamber (CDC),
an array of aerogel threshold \v{C}erenkov counters (ACC),
time-of-flight (TOF) scintillation counters,
and an electromagnetic calorimeter comprised of CsI(Tl) crystals (ECL)
located inside a superconducting solenoid coil
that provides a 1.5~T magnetic field.
An iron flux-return located outside of the coil is instrumented
to detect $\kl$ mesons and to identify muons (KLM).

We reconstruct $\bz$ decays to the following $CP$ 
eigenstates~\cite{footnote:cc}:
$\jpsi\ks$, $\psi(2S)\ks$, $\chi_{c1}\ks$, $\eta_c\ks$ for $\xi_f = -1$
and $\jpsi\kl$ for $\xi_f = +1$.
We also use $\bz \to \jpsi\kstarz$ decays where $\kstarz \to \ks\piz$.
Here the final state is a mixture of even and odd $CP$,
depending on the relative orbital angular momentum of the $\jpsi$ and $\kstarz$.
We find that the final state is primarily $\xi_f = +1$;
the $\xi_f = -1$ fraction is
$0.19 \pm 0.02 \textrm{(stat)} \pm 0.03 \textrm{(syst)}$~\cite{bib:itoh}.
$\jpsi$ and $\psi(2S)$ mesons are reconstructed
via their decays to $\ell^+\ell^-$ ($\ell = \mu,e$).
The $\psi(2S)$ is also reconstructed via $\jpsi\pip\pim$,
and the $\chi_{c1}$ via $\jpsi\gamma$.
The $\eta_c$ is detected in the $\ks\km\pip$, $\kp\km\piz$,
and $p\overline{p}$ modes.
For the $\jpsi\ks$ mode, we use $\ks \to \pip\pim$ and $\piz\piz$ decays;
for other modes we only use $\ks \to \pip\pim$.
For reconstructed $B \to \fcp$ candidates other than $\jpsi\kl$,
we identify $B$ decays using the energy difference $\dE \equiv \EB - \Ebeam$
and the beam-energy constrained mass $\mb \equiv \sqrt{(\Ebeam)^2-(\pB)^2}$,
where $\Ebeam$ is the beam energy in the center-of-mass system (cms)
of the $\Upsilon(4S)$ resonance, and $\EB$ and $\pB$ are
the cms energy and momentum of the reconstructed $B$ candidate, respectively.
\begin{figure*}
  \includegraphics[width=0.45\textwidth,clip]{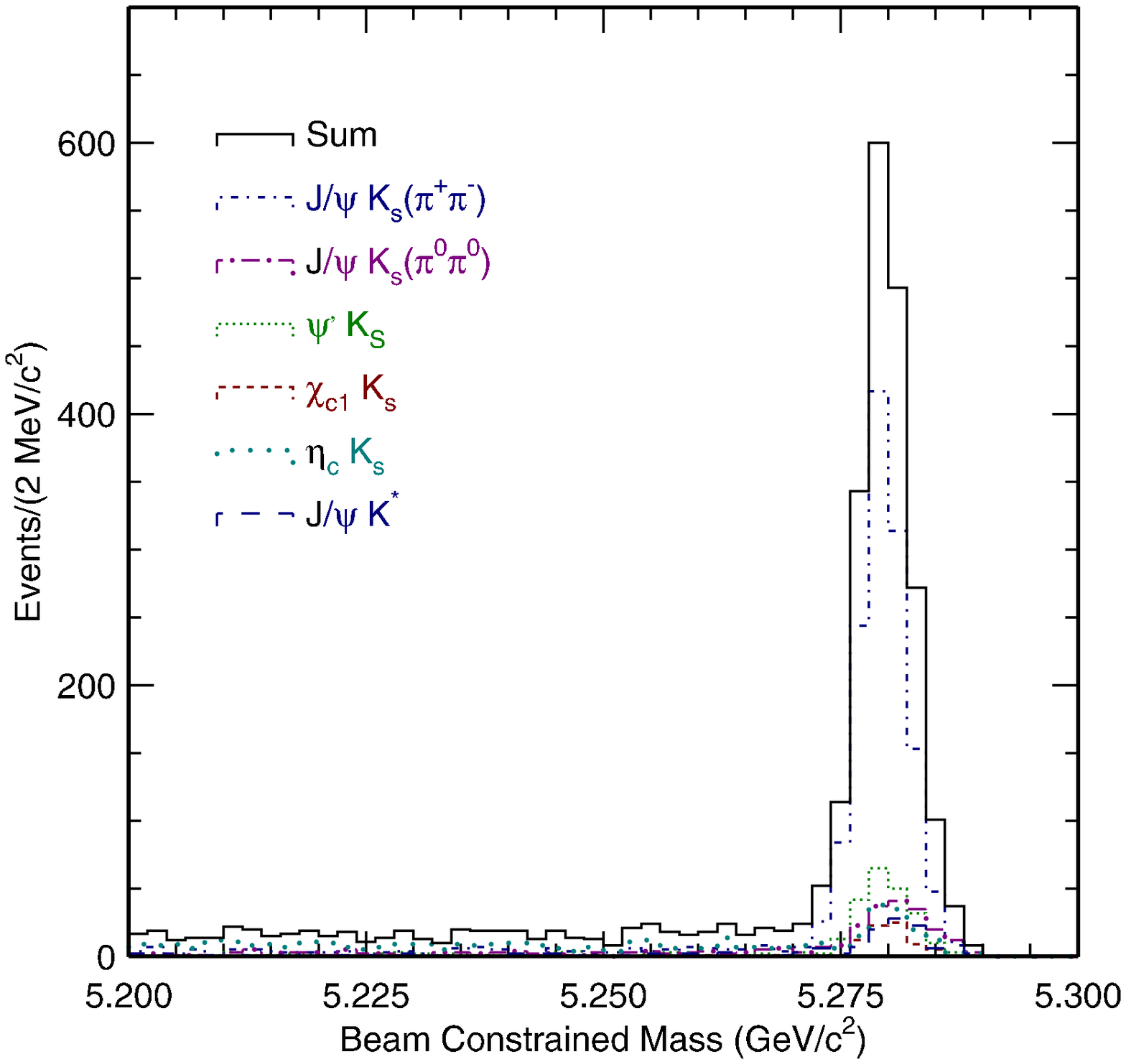}
  \includegraphics[width=0.49\textwidth,clip]{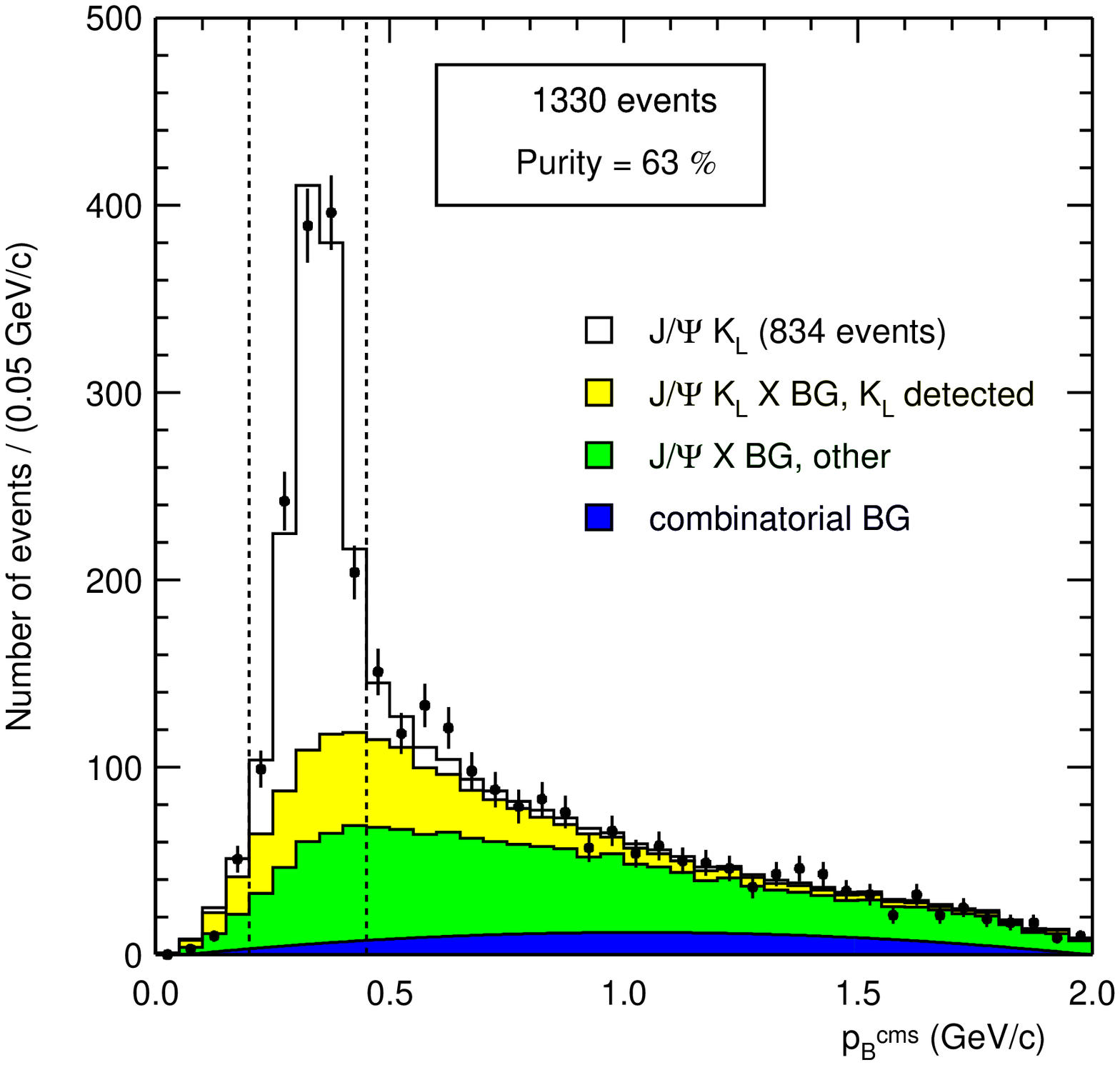}
  \caption{\label{fig:mbc} The beam-energy constrained mass distribution
    for all decay modes other than $\jpsi\kl$ (left).
    The $p_B^\textrm{cms}$ distribution for $\bz \to \jpsi\kl$ candidates
    with the results of the fit (right).}
\end{figure*}
\begin{table*}
  \caption{\label{tab:number} The numbers of reconstructed $B \to \fcp$
    candidates before flavor tagging and vertex reconstruction, $\Nrec$,
    the numbers of events used for the $\sinbb$ determination, $\Nev$,
    and the estimated signal purity in the signal region for each $\fcp$ mode.}
  \begin{ruledtabular}
    \begin{tabular}{lcrrl}
      Mode & $\xi_f$ & $\Nrec$ & $\Nev$ & \multicolumn{1}{c}{Purity} \\
      \hline
      $\jpsi(\ell^+\ell^-)\ks(\pip\pim)$     & $-1$ & 1285 & 1116 & $0.976 \pm 0.001$ \\
      $\jpsi(\ell^+\ell^-)\ks(\piz\piz)$     & $-1$ &  188 &  162 & $0.82 \pm 0.02$ \\
      $\psi(2S)(\ell^+\ell^-)\ks(\pip\pim)$  & $-1$ &   91 &   76 & $0.96 \pm 0.01$ \\
      $\psi(2S)(\jpsi\pip\pim)\ks(\pip\pim)$ & $-1$ &  112 &   96 & $0.91 \pm 0.01$ \\
      $\chi_{c1}(\jpsi\gamma)\ks(\pip\pim)$  & $-1$ &   77 &   67 & $0.96 \pm 0.01$ \\
      $\eta_c(\ks\km\pip)\ks(\pip\pim)$      & $-1$ &   72 &   63 & $0.65 \pm 0.04$ \\
      $\eta_c(\kp\km\piz)\ks(\pip\pim)$      & $-1$ &   49 &   44 & $0.72 \pm 0.04$ \\
      $\eta_c(p\overline{p})\ks(\pip\pim)$   & $-1$ &   21 &   15 & $0.94 \pm 0.02$ \\
      \hline
      All with $\xi_f = -1$                  & $-1$ & 1895 & 1639 & $0.936 \pm 0.003$ \\
      \hline
      $\jpsi(\ell^+\ell^-)\kstarz(\ks\piz)$ & $-1$(19\%)/$+1$(81\%) & 101 & 89 & $0.92 \pm 0.01$ \\
      \hline
      $\jpsi(\ell^+\ell^-)\kl$               & $+1$ & 1330 & 1230 & $0.63 \pm 0.04$ \\
      \hline \hline
      \multicolumn{2}{l}{All}                       & 3326 & 2958 & $0.81 \pm 0.01$ \\
    \end{tabular}
  \end{ruledtabular}
\end{table*}
Figure~\ref{fig:mbc} (left) shows the $\mb$ distributions
for all $\bz$ candidates except for $\bz \to \jpsi\kl$
that have $\dE$ values in the signal region.
Table~\ref{tab:number} lists the numbers of observed candidates, $\Nrec$.

Candidate $\bz \to \jpsi\kl$ decays are selected by requiring
ECL and/or KLM hit patterns that are consistent with the presence
of a shower induced by a $\kl$ meson.
The centroid of the shower is required to be within a $45^\circ$ cone
centered on the $\kl$ direction inferred from
two-body decay kinematics and the measured four-momentum of the $\jpsi$.
Figure~\ref{fig:mbc} (right) shows the $\pB$ distribution,
calculated with the $\bz \to \jpsi\kl$ two-body decay hypothesis.
The histograms are the results of a fit
to the signal and background distributions.
There are 1330 entries in total
in the $0.20 \le \pB \le 0.45$~GeV/$c$ signal region;
the fit indicates a signal purity of 63\%.
The reconstruction and selection criteria for all $\fcp$ channels
used in the measurement are described in more detail elsewhere~\cite{bib:cpv}.

Charged leptons, pions, kaons, and $\Lambda$ baryons
that are not associated with a reconstructed $CP$ eigenstate decay
are used to identify the $b$-flavor of the accompanying $B$ meson.
Based on the measured properties of these tracks, two parameters,
$q$ and $r$, are assigned to an event.
The first, $q$, has the discrete value $+1$~($-1$)
when the tag-side $B$ meson is likely to be a $\bz$~($\bzb$),
and the parameter $r$ is an event-by-event Monte-Carlo-determined
flavor-tagging dilution factor that ranges
from $r=0$ for no flavor discrimination
to $r=1$ for an unambiguous flavor assignment.
It is used only to sort data into six intervals of $r$,
according to estimated flavor purity.
The wrong-tag probabilities, $w_l~(l=1,6)$,
that are used in the final fit are determined directly from data.
Samples of $\bz$ decays to exclusively reconstructed self-tagging channels
are utilized to obtain $w_l$ using
time-dependent $\bz$-$\bzb$ mixing oscillation:
$(\Nof-\Nsf)/(\Nof+\Nsf) = (1-2w_l)\cos(\dM\Dt)$,
where $\Nof$ and $\Nsf$ are the numbers of opposite and same flavor events.
The event fractions and wrong tag fractions
are summarized in Table~\ref{tab:wtag}.
\begin{table}
  \caption{\label{tab:wtag} The event fractions $\epsilon_l$,
    wrong tag fractions $w_l$, and effective tagging efficiencies
    $\eeff^l = \epsilon_l(1-2w_l)^2$ for each $r$ interval.
    The errors include both statistical and systematic uncertainties.
    The event fractions are obtained from the $\jpsi\ks$ simulation.}
  \begin{ruledtabular}
    \begin{tabular}{cccll}
      $l$ & $r$ interval & $\epsilon_l$ &\multicolumn{1}{c}{$w_l$} & \multicolumn{1}{c}{$\eeff^l$} \\
      \hline
      1 & 0.000 -- 0.250 & 0.398 & $0.458\pm0.006$ & $0.003\pm0.001$ \\
      2 & 0.250 -- 0.500 & 0.146 & $0.336\pm0.009$ & $0.016\pm0.002$ \\
      3 & 0.500 -- 0.625 & 0.104 & $0.228\pm0.010$ & $0.031\pm0.002$ \\
      4 & 0.625 -- 0.750 & 0.122 & $0.160~^{+0.009}_{-0.008}$ & $0.056\pm0.003$ \\
      5 & 0.750 -- 0.875 & 0.094 & $0.112\pm0.009$ & $0.056\pm0.003$ \\
      6 & 0.875 -- 1.000 & 0.136 & $0.020\pm0.006$ & $0.126~^{+0.003}_{-0.004}$ \\
    \end{tabular}
  \end{ruledtabular}
\end{table}
The total effective tagging efficiency is determined to be
$\eeff \equiv \sum_{l=1}^6 \epsilon_l(1-2w_l)^2 = 0.288 \pm 0.006$,
where $\epsilon_l$ is the event fraction for each $r$ interval.
The error includes both statistical and systematic uncertainties.
Improvements in the Monte Carlo simulation~\cite{bib:kakuno}
and in the track reconstruction yield $\eeff$
that is higher by 6.7\% (relative) than the value in Ref.~\cite{bib:cpv}.

The vertex position for the $\fcp$ decay is reconstructed
using leptons from $\jpsi$ decays or charged hadrons from $\eta_c$ decays,
and that for $\ftag$ is obtained with well reconstructed tracks
that are not assigned to $\fcp$.
Tracks that are consistent with coming from a $\ks\to\pip\pim$ decay
are not used.
Each vertex position is required to be consistent with
the interaction region profile, determined run-by-run,
smeared in the $r$-$\phi$ plane to account for the $B$ meson decay length.
With these requirements, we are able to determine a vertex
even with a single track;
the fraction of single-track vertices is about 10\% for $\zcp$
and 22\% for $\ztag$.
The proper-time interval resolution function $\Rsig(\Dt)$
is formed by convolving four components:
the detector resolutions for $\zcp$ and $\ztag$,
the shift in the $\ztag$ vertex position
due to secondary tracks originating from charmed particle decays,
and the kinematic approximation that the $B$ mesons are
at rest in the cms~\cite{bib:blife}.
A small component of broad outliers in the $\Dz$ distribution,
caused by mis-reconstruction, is represented by a Gaussian function.
We determine twelve resolution parameters from fits of data
to the neutral and charged $B$ meson lifetimes~\cite{bib:blife}
and obtain an average $\Dt$ resolution of $\sim 1.43$~ps (rms).
The width of the outlier component
is determined to be $42^{+5}_{-4}$~ps;
the fractions of the outlier components are $(2 \pm 1) \times 10^{-4}$
for events with both vertices reconstructed with more than one track,
and $(2.7 \pm 0.2) \times 10^{-2}$ for events with at least
one single-track vertex.

\begin{figure}
  \includegraphics[width=0.6\textwidth,clip]{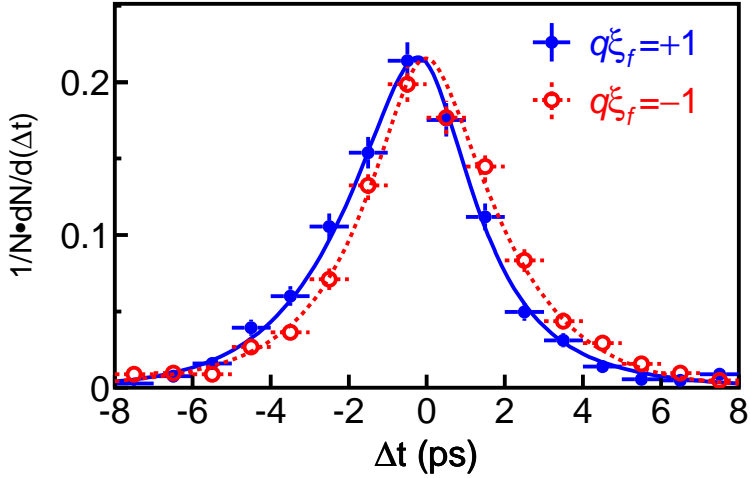}
  \caption{\label{fig:cpfit} The $\Dt$ distributions for the events
    with $q\xi_f = +1$ (solid points) and $q\xi_f = -1$ (open points).
    The results of the global fit with $\sinbb=0.719$ are shown
    as solid and dashed curves, respectively.}
\end{figure}

After flavor tagging and vertexing,
we find 1465 events with $q = +1$ flavor tags
and 1493 events with $q = -1$.
Table~\ref{tab:number} lists the numbers of candidates
used for the $\sinbb$ determination, $\Nev$,
and the estimated signal purity in the signal region
for each $\fcp$ mode.
Figure~\ref{fig:cpfit} shows the observed $\Dt$ distributions
for the $q\xi_f = +1$ (solid points)
and $q\xi_f = -1$ (open points) event samples.
The asymmetry between the two distributions demonstrates
the violation of $CP$ symmetry.
We determine $\sinbb$ from an unbinned maximum-likelihood fit
to the observed $\Dt$ distributions.
The probability density function (PDF) expected
for the signal distribution is given by
\begin{equation}
  \label{eq:deltat}
  \Psig(\Dt, q, w_l, \xi_f) =
  \frac{e^{-|\Dt|/\taubz}}{4\taubz}[1 - q\xi_f(1-2w_l)\sinbb\sin(\dM\Dt)],
\end{equation}
where we fix the $\bz$ lifetime $\taubz$ and mass difference
at their world average values~\cite{bib:pdg}.
Each PDF is convolved with the appropriate $\Rsig(\Dt)$
to determine the likelihood value for each event as a function of $\sinbb$:
\begin{eqnarray}
  P_i &=& (1-\fol) \int \Bigl[ \fsig \Psig(\Dt',q,w_l,\xi_f)\Rsig(\Dt-\Dt')
  \nonumber \\
  && \quad + \; (1-\fsig)\Pbkg(\Dt')\Rbkg(\Dt-\Dt')\Bigr] d\Dt'
  + \fol \Pol(\Dt),
\end{eqnarray}
where $\fsig$ is the signal fraction
calculated as a function of $\pB$ for $\jpsi\kl$
and of $\dE$ and $\mb$ for other modes.
$\Pbkg(\Dt)$ is the PDF for combinatorial background events,
which is modeled as a sum of exponential and prompt components.
It is convolved with a sum of two Gaussians, $\Rbkg$,
which is regarded as a resolution function for the background.
To account for a small number of events that give large $\Dt$ in
both the signal and background, we introduce
the PDF of the outlier component, $\Pol$, and its fraction $\fol$.
The only free parameter in the final fit is $\sinbb$,
which is determined by maximizing the likelihood function $L = \prod_i P_i$,
where the product is over all events.
The result of the fit is
\[
\sinbb = 0.719 \pm 0.074 \textrm{(stat)} \pm 0.035 \textrm{(syst)} .
\]
The systematic error is dominated by uncertainties
in the vertex reconstruction (0.022).
Other significant contributions come from uncertainties in $w_l$ (0.015),
the resolution function parameters (0.014),
a possible bias in the $\sinbb$ fit (0.011),
and the $\jpsi\kl$ background fraction (0.010).
The errors introduced by uncertainties in $\dM$ and $\taubz$
are less than 0.010.

Several checks on the measurement are performed.
Table~\ref{tab:check} lists the results obtained by applying the same analysis
to various subsamples.
\begin{table}
  \caption{\label{tab:check} The numbers of candidate events, $\Nev$,
    and values of $\sinbb$ for various subsamples (statistical errors only).}
  \begin{ruledtabular}
    \begin{tabular}{lrc}
      Sample & \multicolumn{1}{c}{$\Nev$} & $\sinbb$ \\
      \hline
      $\jpsi\ks(\pip\pim)$    & 1116                         & $0.73\pm0.10$ \\
      $(c\overline{c})\ks$ except $\jpsi\ks(\pip\pim)$ & 523 & $0.67\pm0.17$ \\
      $\jpsi\kl$              & 1230                         & $0.78\pm0.17$ \\
      $\jpsi\kstarz(\ks\piz)$ &   89                         & $0.04\pm0.63$ \\
      \hline
      $\ftag = \bz$ ($q=+1$)  & 1465                         & $0.65\pm0.12$ \\
      $\ftag = \bzb$ ($q=-1$) & 1493                         & $0.77\pm0.09$ \\
      \hline
      $0 < r \le 0.5$         & 1600                         & $1.27\pm0.36$ \\
      $0.5 < r \le 0.75$      &  658                         & $0.62\pm0.15$ \\
      $0.75 < r \le 1$        &  700                         & $0.72\pm0.09$\\
      \hline
      data before 2002        & 1587                         & $0.78\pm0.10$ \\
      data in 2002            & 1371                         & $0.65\pm0.11$ \\
      \hline \hline
      All                     & 2958                         & $0.72\pm0.07$ \\
    \end{tabular}
  \end{ruledtabular}
\end{table}
All values are statistically consistent with each other.
\begin{figure}
  \includegraphics[width=0.8\textwidth,clip]{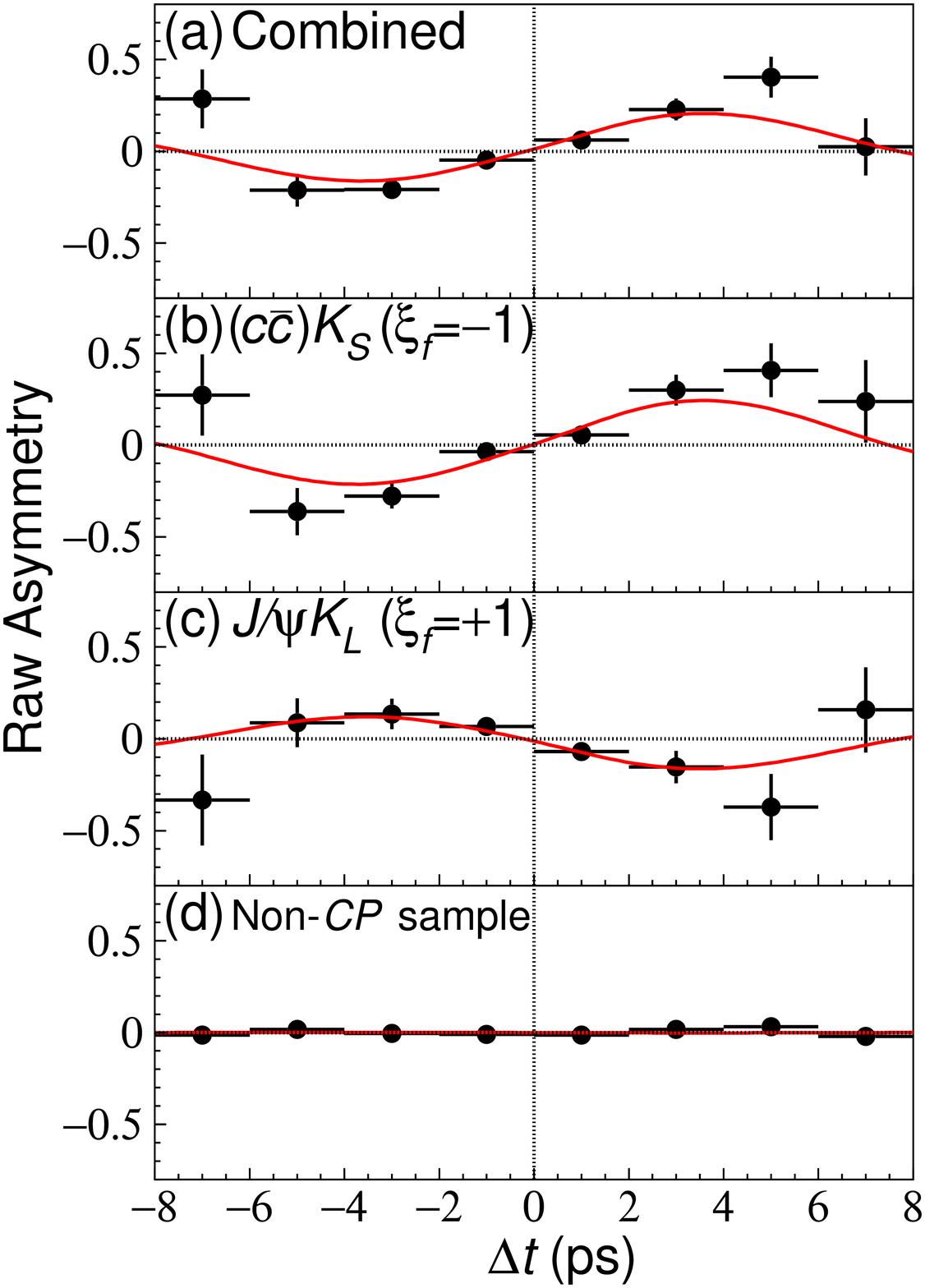}
  \caption{\label{fig:rawasym} (a) The raw asymmetry for all modes combined.
    The asymmetry for $\jpsi\kl$ and $\jpsi\kstarz$ is inverted
    to account for the opposite $CP$ eigenvalue.
    The corresponding plots for (b) $(c\overline{c})\ks$,
    (c) $\jpsi\kl$, and (d) non-$CP$ control samples are also shown.
    The curves are the results of the unbinned maximum likelihood fit
    applied separately to the individual data samples.}
\end{figure}
Figures~\ref{fig:rawasym}(a)--\ref{fig:rawasym}(c) show the raw asymmetries
and the fit results for all modes combined, $(c\overline{c})\ks$,
and $\jpsi\kl$, respectively.
A fit to the non-$CP$ eigenstate modes $\bz \to D^{(*)-}\pip$,
$D^{*-}\rho^+$, $\jpsi\kstarz(\kp\pim)$, and $D^{*-}\ell^+\nu$,
where no asymmetry is expected,
yields $0.005 \pm 0.015$(stat).
Figure~\ref{fig:rawasym}(d) shows the raw asymmetry
for these non-$CP$ control samples.

The signal PDF for a neutral $B$ meson decaying into a $CP$ eigenstate
[Eq.~(\ref{eq:deltat})] can be expressed in a more general form as
\begin{align}
  \label{eq:deltat_general}
  \Psig(\Dt, q, w_l) =
  \frac{ e^{-|\Dt|/\taubz} }{4\taubz}
  \Biggl\{ 1 + q(1-2w_l)
  & \left[ \frac{2 \textrm{Im}\lambda}{|\lambda|^2 + 1} \sin(\dM\Dt) \right. \nonumber \\
  & \left. \quad + \frac{|\lambda|^2 - 1}{|\lambda|^2 + 1} \cos(\dM\Dt) \right] \Biggr\} ,
\end{align}
where $\lambda$ is a complex parameter that depends on both
$\bz$-$\bzb$ mixing and on the amplitudes for $\bz$ and $\bzb$ decay
to a $CP$ eigenstate.
The presence of the cosine term ($|\lambda| \neq 1$)
would indicate direct $CP$ violation;
the value for $\sinbb$ reported above is determined
with the assumption $|\lambda| = 1$, as $|\lambda|$ is expected
to be very close to one in the SM.
In order to test this assumption,
we also performed a fit using the above expression with
$a_{CP} = -\xi_f \text{Im}\lambda/|\lambda|$
and $|\lambda|$ as free parameters, keeping everything else the same.
We obtain
\[
|\lambda| = 0.950 \pm 0.049 \textrm{(stat)} \pm 0.025 \textrm{(syst)}
\]
and $a_{CP} = 0.720 \pm 0.074 \textrm{(stat)}$
for all $CP$ modes combined, where the sources of the systematic error
for $|\lambda|$ are the same as those for $\sinbb$.
This result is consistent with the assumption used in our analysis.

\begin{acknowledgments}
We wish to thank the KEKB accelerator group for the excellent
operation of the KEKB accelerator.
We acknowledge support from the Ministry of Education,
Culture, Sports, Science, and Technology of Japan
and the Japan Society for the Promotion of Science;
the Australian Research Council
and the Australian Department of Industry, Science and Resources;
the National Science Foundation of China under contract No.~10175071;
the Department of Science and Technology of India;
the BK21 program of the Ministry of Education of Korea
and the CHEP SRC program of the Korea Science and Engineering Foundation;
the Polish State Committee for Scientific Research
under contract No.~2P03B 17017;
the Ministry of Science and Technology of the Russian Federation;
the Ministry of Education, Science and Sport of the Republic of Slovenia;
the National Science Council and the Ministry of Education of Taiwan;
and the U.S.\ Department of Energy.
\end{acknowledgments}



\begin{thebibliography}{99}

\bibitem{bib:ckm}
  M.~Kobayashi and T.~Maskawa, Prog. Theor. Phys. \textbf{49}, 652 (1973).

\bibitem{bib:sanda}
  A.~B.~Carter and A.~I.~Sanda, Phys. Rev. D \textbf{23}, 1567 (1981);
  I.~I.~Bigi and A.~I.~Sanda, Nucl. Phys. \textbf{B193}, 85 (1981).

\bibitem{bib:cpv}
  Belle Collaboration, K.~Abe \textit{et al.}, Phys. Rev. Lett. \textbf{87}, 091802 (2001);
  Phys. Rev. D \textbf{66}, 032007 (2002).

\bibitem{bib:babar}
  BaBar Collaboration, B.~Aubert \textit{et al.}, Phys. Rev. Lett. \textbf{87}, 091801 (2001);
  Phys. Rev. D \textbf{66}, 032003 (2002).

\bibitem{bib:blife}
  Belle Collaboration, K.~Abe \textit{et al.}, Phys. Rev. Lett. \textbf{88}, 171801 (2002).

\bibitem{bib:belle}
  Belle Collaboration, A.~Abashian \textit{et al.},
  Nucl. Instrum. Methods Phys. Res. A \textbf{479}, 117 (2002).

\bibitem{bib:kekb}
  KEK Report No. 2001-157, edited by E.~Kikutani, 2001
  [Nucl. Instrum. Methods Phys. Res. A (to be published)].

\bibitem{footnote:cc}
  Throughout this paper, when a decay mode is quoted,
  the inclusion of the charge conjugate mode is implied.

\bibitem{bib:itoh}
  Belle Collaboration, K.~Abe \textit{et al.}, Phys. Lett. B \textbf{538}, 11 (2002).

\bibitem{bib:kakuno}
  H.~Kakuno \textit{et al.}, (unpublished).

\bibitem{bib:pdg}
  Particle Data Group, K.~Hagiwara \textit{et al.}, Phys. Rev. D \textbf{66}, 010001 (2002).

\end{thebibliography}
\end{document}